\documentclass[12pt]{article}

\usepackage[dvips]{graphicx}

\newcommand{\nn}{\nonumber}
\newcommand{\be}{\begin{equation}}
\newcommand{\ee}{\end{equation}}
\newcommand{\ba}{\begin{eqnarray}}
\newcommand{\ea}{\end{eqnarray}}
\newcommand{\ci}[1]{\cite{#1}}
\def\={\,=\,}

\newcommand{\LQCD}{\Lambda_{\rm{QCD}}}
\def\als{\alpha_s}

\def\mev{\,{\rm MeV}}
\def\gev{\,{\rm GeV}}

\newcommand{\da}{{distribution amplitude}}

\def\muR{\mu_R}
\def\muF{\mu_F}
\def\taub{\bar{\tau}}
\newcommand{\tw}{\textwidth}                          
\newcommand{\req}[1]{(\ref{#1})}

\def\sh{\hat{s}}
\def\uh{\hat{u}}
\def\th{\hat{t}}

\begin{document}
\thispagestyle{empty}
\begin{flushright}
\today\\[20mm]
\end{flushright}

\begin{center}

  {\Large\bf Wide-angle pion-$\Delta(1232)$ photoproduction} \\[0.3em]
\vskip 15mm
P.\ Kroll \\[1em]
{\small {\it Fachbereich Physik, Universit\"at Wuppertal, D-42097 Wuppertal,
Germany}}\\

\end{center}

\begin{abstract}
  High-energy wide-angle photoproduction of $\pi\Delta$ final states is investigated within the handbag
  mechanism to twist-3 accuracy. In this approach the process amplitudes factorize in a hard
  partonic subprocess and in form factors which represent $1/x$-moments of $p-\Delta$ transition
  generalized parton distributions (GPDs) at zero skewness. The subprocess, calculated to twist-3
  accuracy, is the same as in  pion photoproduction. The $p-\Delta$ GPDs are related to the
  proton-proton GPDs exploiting large-$N_C$ results. The proton-proton GPDs as well as the twist-2 and
  twist-3 pion distribution amplitudes appearing in the subprocess are taken from other work. Reasonable
  agreement with experiment is found in this almost parameter-free analysis.
\end{abstract}  

%%%%%%%%%%%%%%%%%%%%%%%%%%%%%%%%%%%%%%%%%%%%%%%%%%%%%%%%%%%%%%%%%%
\section{Introduction}
%%%%%%%%%%%%%%%%%%%%%%%%%%%%%%%%%%%%%%%%%%%%%%%%%%%%%%%%%%%%%%%%%%
Factorization properties of QCD allow to calculate inclusive and exclusive
scattering processes. For exclusive processes, in particular, there are two 
kinematical regions in which factorization properties hold. On the one hand this is
the generalized Bjoerken region of large photon virtualities, $Q^2$, large energies
in the photon-nucleon center-of-mass system (c.m.s.) and fixed Bjorken-$x$ but small
Mandelstam $-t$ ($-t\ll Q^2$). In this kinematical region it has been shown \ci{ji,collins}
that the process amplitudes are represented as a convolution of hard perturbatively
calculable partonic subprocess and soft hadronic matrix elements parameterized as GPDs.
There are many applications of this theoretical concept as for instance deeply virtual
Compton scattering or meson production, see for instance the reviews \ci{belitsky,diehl03}.
The other kinematical region in which factorization applies, is the wide-angle region, i.e.\
the region where all three Mandelstam variables, $s$, $-t$ and $-u$, are large. Here the
process amplitudes factorize in hard pertonic subprocesses and soft form factors
representing $1/x$-moments of GPDs. This so-called handbag mechanism has been applied
to wide-angle Compton scattering \ci{radyushkin98,DFJK1} and photoproduction of pseudoscalar
mesons \ci{HK00,KPK18}. It also applies to electroproduction of mesons provided $Q^2\ll -t$
\ci{KPK21}. Lack of data prevented further applications as for instance photoproduction of
vector mesons.

Both these classes of hard exclusive processes, the deeply virtual as well as the wide-angle ones,
are in a sense complementary. In the deeply-virtual region the GPDs are probed at small $-t$,
typically less than $1\,\gev^2$ while in the wide-angle region the GPDs are examined at large $-t$.
The knowledge of the large $-t$ behavior of the GPDs is important for studying the impact parameter
distribution of partons inside the proton. As shown by Burkardt \ci{burkardt03} the impact-parameter
distribution of the partons is obtained from a Fourier transform of zero-skewness GPDs. The impact
parameter is canonically conjugated to the momentum transfer from the initial to the final baryon.
The square of the momentum transfer is the Mandelstem $t$. Obviously, for a reliable Fourier
transform knowledge of the GPDs is required in a fairly large range of $t$, well beyond $1\,\gev^2$.

Recently the interest in transition GPDs has strongly grown both experimentally and theoretically \ci{diehl24}.
Besides baryon octet-octet transitions also octet-decuplet transitions and particularly $p-\Delta$ ones
are of interest. Data relevant for the latter transition GPDs are scarce up to now. In the deeply virtual
region only the beam spin asymmetry has been measured by the CLAS collaboration \ci{diehl23}.
There is also a theoretical study of the $\pi^-\Delta^{++}(1232)$ electroproduction  cross section in the
generalized Bjorken regime \ci{KPK22}. The $p-\Delta$ transition GPDs are fixed in the large-$N_C$ limit
where relations between the $p-\Delta$ and the proton-proton GPDs hold \ci{belitsky,frankfurt99}.
Here, in this article, wide-angle photoproduction of $\pi\Delta$ final states will be investigated.
This process has been measured in the wide-angle region at high energies at SLAC \ci{anderson} long ago.
The calculation of wide-angle $\pi\Delta$ photoproduction goes along the same line as pion photoproduction
\ci{KPK18}: the partonic subprocess is identical and has been calculated in \ci{KPK18} to twist-3 accuracy
and leading-order of perturbative QCD. The transition GPDs are fixed in the large-$N_C$ limit as in \ci{KPK22}.
Thus, the $\pi\Delta$ photoproduction cross section can be calculated almost free of parameters and as will
be shown in the following, reasonable agreement with experiment is found.

The plan of the paper is as follows: In the next section the handbag mechanism for wide-angle photoproduction
will be introduced. In Sect. 3 the subprocess and in Sect. 4 the GPDs in the large-$N_C$
limit will be discussed. The results for the photoproduction cross section will be presented in Sect. 5
and compared to experiment. The paper will end with a summary.

%%%%%%%%%%%%%%%%%%%%%%%%%%%%%%%%%%%%%%%%%%%%%%%%%%%%%%%%%%%%%%%%%%%%
\section{The handbag mechanism for wide-angle photoproduction}
%%%%%%%%%%%%%%%%%%%%%%%%%%%%%%%%%%%%%%%%%%%%%%%%%%%%%%%%%%%%%%%%%%%%
Here, in this article, the process
\be
\gamma(q,\mu)\;p(p,\nu)\to \pi(q',0)\;\Delta(p',\nu')
\label{eq:process}
\ee
will be investigated in the wide-angle region where the three Mandelstam
variables, $s$, $-t$ and $-u$, are much larger than $\Lambda^2$. The constant $\Lambda$
is a typical hadronic scale of order $1\,\gev$. In Eq.\ \req{eq:process}
$q, p, p'$ denote the momenta of the involved particles and $\mu, \nu, \nu'$
their helicities. It is advantageous to work in a symmetric frame in which
the skewness, $\xi$, defined by the following ratio of light-cone plus components
of the baryon momenta
\be
\xi\=\frac{(p-p')^+}{(p+p')^+}
\label{eq:skewness}
\ee
is zero. Except of baryon-mass corrections the particle
momenta are parameterized as 
\ba
p&=& \Big[p^+,-\frac{t}{8p^+},-\frac12\sqrt{-t},0\Big]\,, \nn\\
p'&=& \Big[p^+,-\frac{t}{8p^+},\phantom{-}\frac12\sqrt{-t},0\Big]\,, \nn\\
q&=& \Big[ q^+,-\frac{t}{8q^+},\phantom{-}\frac12\sqrt{-t},0\Big]\,,
\ea
in light-cone coordinates  where
\be
p^+\=\big(\sqrt{s}+\sqrt{-u}\big)/4\,, \qquad q^+\=\big(\sqrt{s}-\sqrt{-u}\big)/4\,.
\ee
The pion momentum is fixed by momentum conservation.

\begin{figure}[t]
  \begin{center}
    \includegraphics[width=0.5\tw]{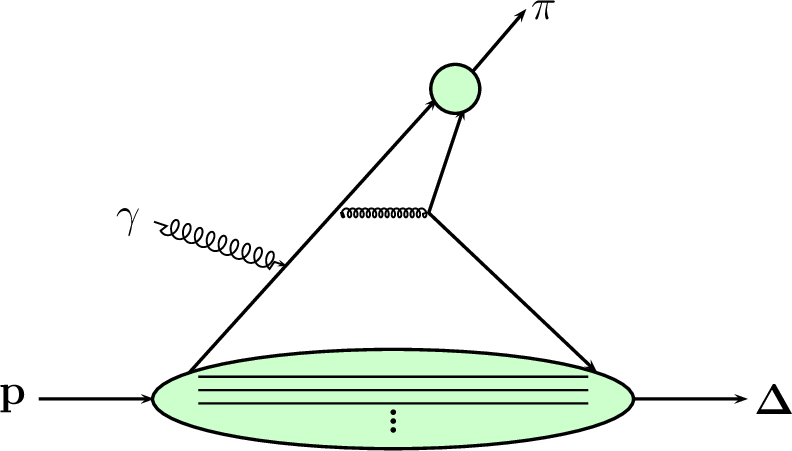}
\end{center}
  \caption{A typical graph for the handbag mechanism. The shaded regions are soft regions
    in which no hard scale appear.}
  \label{fig:handbag}
  \end{figure}
In the handbag mechanism a (light-cone) helicity amplitude, ${\cal M}_{0\nu',\mu\nu}$,
for the process \req{eq:process} factorize in a product of subprocess
amplitudes, ${\cal H}$, for pion photoproduction off quarks and form factors which represent
the soft physics that controls the emission and reabsorption of quarks from the baryons, see
Fig.\ \ref{fig:handbag}.
These form factors represent $1/x$-moments of zero-skewness proton-$\Delta$ GPDs.
The arguments for this type of factorization are presented  in \ci{DFJK1,HK00} in  detail.
Thus, a brief repetition of the arguments should be enough:
It is assumed that the parton virtualities are restricted by $k_i^2<\Lambda^2$ and that the
intrinsic transverse momenta, $k_{\perp i}$, defined with respect to their parent hadron's
momentum, satisfy the condition $k_{\perp i}/x_i<\Lambda^2$ where $x_i$ is the momentum fraction
that parton $i$ carries. One can then show that, up to corrections of order $\Lambda/\sqrt{-t}$, the
subprocess Mandelstam variables, $\sh, \th, \uh$, coincide with the ones for the full process
\be
\th=t\,, \quad \sh\=(k_j+q)^2\simeq (p+q)^2\=s\,, \quad \uh\=(k_j'-q)^2\simeq (p'-q)^2\=u\,,
\label{eq:assignment}
\ee
where $k_j$ and $k_j'=k_j+q-q'$ denote the momenta of the active partons, i.e. those partons to
which the photon couple. Hence, the active partons are approximately on-shell, move collinear
with their parent hadrons and carry momentum fractions close to unity, $x_j,x_j'\simeq 1$.
As in deeply virtual exclusive scattering , the physical situation is that of a hard parton-level
subprocess, $\gamma q_a\to \pi q_b$, and a soft emission and reabsorption of quarks from the baryons.
Thus, up to corrections of order $\Lambda/\sqrt{-t}$, one can write a helicity amplitude for
$\gamma p\to \pi \Delta$ as
\be
   {\cal M}_{0\nu',\mu\nu}\= e_0\int \frac{dx}{x} \sum_{\lambda',\lambda} {\cal H}_{0\lambda',\mu\lambda}
   A_{\nu'\lambda',\nu\lambda}
   \label{eq:amplitudes}
   \ee
   where $\lambda$ and $\lambda'$ denote the  helicities of the emitted and reabsorbed quark
   and $e_0$ is the positron charge. The soft $p-\Delta$ matrix elements, $A$, of quark field
   operators are \ci{KPK22}
   \be
   A_{\nu'\lambda',\nu\lambda}\=\int \frac{z^-}{2\pi} e^{ixP^+z^-}\langle \Delta^{++}(p',\nu')|
        {\cal O}_{\lambda'\lambda}|p(p,\nu)\rangle|_{z^+=0,z_\perp=0}
   \ee
   where
  \ba
     {\cal O}_{\lambda\lambda}&=& \frac14\bar{u}(-z/2)\gamma^+(1+2\lambda \gamma_5)d(z/2)\,, \nn\\
     {\cal O}_{-\lambda\lambda}&=& -\frac{i\lambda}{2}\bar{u}(-z/2)
                                \big(\sigma^{+1}-2\lambda i\sigma^{+2}\big)d(z/2)\,.
   \ea
   These matrix elements describe the emission an of on-shell quarks with helicity $\lambda$ and
   reabsorption of an on-shell quark with helicity $\lambda'$ \ci{diehl01}. They are parameterized
   as $p-\Delta$ transition GPDs \ci{belitsky,KPK22} and are explicitly given in
   \ci{KPK22}. For wide-angle photoproduction  only the matrix elements at zero skewness are needed.
   Nevertheless 14 of the altogether 16 $p-\Delta$ GPDs contribute at this value of skewness. Only 
   the  GPDs $G_4$ and $\tilde{G}_4$ and their associated form factors decouple.   

For the process of interest there are three charge configurations in the final state:
$\pi^-\Delta^{++}$, $\pi^0\Delta^+$ and  $\pi^+\Delta^0$. The corresponding GPDs and their
associated form factors are related by isospin symmetry \ci{belitsky}
\be
G^{p\Delta^{++}}(x,t) \= -\frac{\sqrt{3}}{2}\, G^{p\Delta^+}(x,t)
                           \=-\sqrt{3}\, G^{p\Delta^0}(x,t)\,.
\label{eq:isospin}
\ee
The skewness variable is omitted in the GPDs for convenience.  One can also generalize the handbag
mechanism to wide-angle electroproduction of $\pi\Delta$ final states provided that the photon
virtuality, $Q^2$, is less than $-t$ and $-u$. This calculation is analogous to the case of
$\pi$-nucleon, see \ci{KPK21}. The photon virtuality affects only the subprocess amplitudes. 
The photoproduction of $\eta\Delta^+$ and $\eta'\Delta^+$ is a direct generalization of the
calculation of $\eta, \eta'$-photoproduction \ci{KPK21a}. Kaon-$\Sigma^*$ production requires
$p-\Sigma^*$ GPDs (with $sd$ flavor content) which are related to the $p-\Delta$ GPDs
(with $du$ flavor content) by SU(3) flavor symmetry \ci{belitsky},
e.g.\
\be
G^{sd}_{p\Sigma^{*0}}\=\frac1{\sqrt{2}}\,G^{du}_{p\Delta^0}\,.
\ee
Otherwise the calculation of these processes is similar to the $\pi\Delta$ case investigated here.
Finally, photoproduction of vector mesons at wide-angles can also be treated with the handbag
mechanism \ci{HK00}. However, for flavor-neutral vector mesons there is an additional gluonic
subprocess, $\gamma g\to Vg$, which goes along with gluonic $p-\Delta$ transition GPDs and
their corresponding form factors.

%%%%%%%%%%%%%%%%%%%%%%%%%%%%%%%%%%%%%%%%%%%%%%%%%%%%%%%%%%%%%%%%%%%%%
\section{The subprocess}
%%%%%%%%%%%%%%%%%%%%%%%%%%%%%%%%%%%%%%%%%%%%%%%%%%%%%%%%%%%%%%%%%%%
The amplitudes for the subprocess $\gamma q_a\to \pi q_b$ have been calculated 
in collinear factorization to twist-3 accuracy~\footnote{
      In the calculation of the subprocess amplitude as well as in the definition of the $p-\Delta$
      matrix elements light-cone gauge is used.}.
At leading-order of perturbative QCD the twist-2 contribution reads \ci{HK00,KPK18}
\ba
       {\cal H}_{0\lambda\mu\lambda}&=& \sqrt{2}\pi\als(\muR){\cal C}_\pi^{(ab)} \frac{C_F}{N_C}
                                   \frac{f_\pi}{\sqrt{-\th}}
                                   \int_0^1\frac{d\tau}{\tau}\,\Phi_\pi(\tau,\muR) \nn\\
                         &\times& \Big[(1+2\lambda\mu)\sh - (1-2\lambda\mu)\uh\Big]\,
                                   \Big(\frac{e_a}{\sh}+\frac{e_b}{\uh}\Big)
\label{eq:twist2}
 \ea
    where the flavor weight factors are
    \be
       {\cal C}^{ud}_{\pi^+}\={\cal C}^{du}_{\pi^-}\=1\,, \qquad
       {\cal C}^{uu}_{\pi^0}\=-{\cal C}^{dd}_{\pi^0}\=\frac1{\sqrt{2}}\,.
    \ee
    All other ${\cal C}^{ab}_\pi$ are zero. The strong coupling constant is evaluated in the
    one-loop approximation using the renormalization (and factorization) scale
    \be
    \mu_R^2\=\mu_F^2\= \frac{\th\uh}{\sh}\,.
    \ee
    This scale takes care of the requirement that all Mandelstam variables should be large.
    Up to mass corrections $\mu_R^2\simeq s/4$ for a c.m.s.  scattering angle near $90^\circ$.
    For the familiar pion decay constant, $f_\pi$, the value $132 \mev$ is taken; $N_C (=3)$ is the
    number of colors and $C_F=(N_C^2-1)/(2N_C)$. The quark charges, $e_a$ and $e_b$, are given
    in units of the positron charge. Finally, $\phi_\pi$ is the twist-2 pion distribution
    amplitude for which the truncated Gegenbauer expansion is used
    \be
    \phi_\pi(\tau,\mu_R)\=6\tau\taub \,\Big[1 +a_2(\mu_0)L^{\gamma_2/\beta_0} C_2^{3/2}(2\tau-1)\Big]
    \label{eq:tw2-DA}
    \ee
    with the lattice QCD result for the second Gegenbauer coefficient \ci{braun15}
    \be
    a_2(\mu_0)\= 0.1364 
    \ee
    quoted at the initial scale $\mu_0=2\,\gev$. The anomalous dimension is $\gamma_2=50/9$
    and $\beta_0=(11N_C-2n_f)/3$ ($n_f=4$). The quantity $L$ is defined as
    \be
    L\=\frac{\alpha_s(\mu_R)}{\alpha_s(\mu_0)}\=\frac{\ln{(\mu_0^2/\LQCD^2)}}{\ln{(\mu_R^2/\LQCD^2)}}
    \ee  
    where the value $\LQCD=0.22\,\gev$ is adopted. The $\tau$-integration in \req{eq:twist2}
    can be carried out analytically for the \da{} \req{eq:tw2-DA}:
    \be
    \int_0^1 \frac{d\tau}{\tau} \phi_\pi(\tau,\muR)\=3\big[1+a_2(\muR)\big]\,.
    \ee

    The twist-3 contribution has been calculated in \ci{KPK18,KPK21} and for details of this
    contribution it is referred to these papers. It consists of two parts - the 2-body one
    (the twist-3 $q\bar{q}$ Fock component of the pion with the \da s $\phi_{\pi p}$ and
    $\phi_{\pi\sigma}$) and the 3-body one (the $q\bar{q}g$ Fock component with the \da{} $\phi_{3\pi}$).
    Both contributions are related to each other by the equation of motion. In light-cone
    gauge the relations read  
    \ba
    \lefteqn{f_\pi\mu_\pi\Big[\taub \phi_{\pi p}(\tau) -\frac{\taub}{6} \frac{d}{d\tau} \phi_{\pi\sigma}(\tau)
                      -\frac13 \phi_{\pi\sigma}(\tau)\Big] = } \hspace*{0.3\tw}\nn\\
         && 2 f_{3\pi} \int_0^{\taub} \frac{d\tau_g}{\tau_g}\,\phi_{3\pi}(\tau,\taub-\tau_g,\tau_g)\,,\nn\\
    \lefteqn{f_\pi\mu_\pi\Big[\tau \phi_{\pi p}(\tau) +\frac{\tau}{6} \frac{d}{d\tau} \phi_{\pi\sigma}(\tau)
                      -\frac13 \phi_{\pi\sigma}(\tau)\Big] = } \hspace*{0.3\tw} \nn\\
     && 2 f_{3\pi} \int_0^{\tau} \frac{d\tau_g}{\tau_g}\,\phi_{3\pi}(\taub-\tau_g,\taub,\tau_g)\,.
\label{eq:motion}
    \ea
   The parameter $\mu_\pi$ appearing in \req{eq:motion} is defined as
   \be
   \mu_\pi\=\frac{m_\pi^2}{m_u+m_d}\,, 
   \ee
   i.e.\ it is pion mass enhanced by the chiral condensate by means of the divergence of the
   axial-vector current ($m_u$ and $m_d$ are current quark masses). At the initial scale
   the value $\mu_\pi=2\,\gev$ is taken. It evolves as
   \be
   \mu_\pi(\muF)\=L^{-4/\beta_0} \mu_\pi(\mu_0)\,.
   \ee
   Because of the relations \req{eq:motion} the 2-body distribution amplitudes $\phi_{\pi p}$ and
   $\phi_{\pi\sigma}$ are not required explicitly although they are fixed by \req{eq:motion}
   for a given 3-body \da{},  $\phi_{3\pi}$. Thus, the full twist-3 contribution to the subprocess
   amplitudes can solely be expressed  by the 3-body \da{}. As the twist-2 \da{} the 3-body one
   is conventionally normalized to unity and accompanied by the normalization parameter $f_{3\pi}$
   which is scale dependent and evolves as
 \be
 f_{3\pi}(\muR)\=L^{(16/3C_F-1)/\beta_0} f_{3\pi}(\mu_0)\,.
   \ee
      
  The combined 2- and 3-body twist-3 amplitudes read \ci{KPK21}
 \ba
   {\cal H}_{0-\lambda\mu\lambda}&=& 2\sqrt{2}\pi \als(\muR) {\cal C}_\pi^{(ab)}\frac{C_F}{N_C}
               f_{3\pi}(\muR) (2\lambda-\mu)\,\frac{\sqrt{-\uh\sh}}{\sh^2\uh^2} \nn\\
               &\times& \int_0^1 d\tau \int_0^{\taub} \frac{d\tau_g}{\tau_g}
                                          \Phi_{3\pi}(\tau,\taub-\tau_g,\tau_g,\muR) \nn\\
               &\times& \left[ \Big(\frac1{\taub^2}-\frac1{\taub(\taub-\tau_g)}\Big)
                                   \big(e_a\uh^2+e_b\sh^2\big)  
                -  \frac{C_G}{C_F}\,\frac{2\th}{\tau\tau_g}
                                 \,\big(e_a\uh + e_b\sh\big) \right]
      \label{eq:twist3}
        \ea
         where  ($C_A=N_C$)
         \be
         C_G\=C_F-\frac12 C_A\,.
         \ee
         One directly sees from \req{eq:twist3} that~\footnote{
                                 Explicit helicities of the subprocess amplitudes as well as of the
                                 full amplitudes, ${\cal M}$, are labeled by their signs.
                                 Those of the $\Delta(1232)$ are denoted by $2\nu'$.}
         \be
         {\cal H}_{0-++}\= {\cal H}_{0+--}\= 0\,,  \qquad {\cal H}_{0--+} \= -{\cal H}_{0++-}\,.
         \ee
         
         Following \ci{braun90} a truncated conformal expansion is used
         for the 3-body \da{}
         \ba
         \Phi_{3\pi}(\tau_1,\tau_2,\tau_g,\muR)&=& 360 \tau_1\tau_2\tau_g^2 \Big( 1
         + \omega_{10}(\mu_R) \frac12 (7\tau_g-3)\nn\\
         &+& \omega_{20}(\muR) ( 2-4\tau_1\tau_2-8\tau_g+8\tau_g^2)  \nn\\
         &+& \omega_{11}(\muR) (3\tau_1\tau_2 -2\tau_g+3\tau_g^2) \Big)\,.
         \label{eq:3-body-DA}
         \ea
The evolution of the expansion coefficients can be found in \ci{KPK21,braun90}.
Note that the coefficients $\omega_{20}$ and $\omega_{10}$ mix under evolution.
For the \da{} \req{eq:3-body-DA} the integrals in \req{eq:twist3} can be carried out analytically:
\ba
   {\cal H}_{0-\lambda\mu\lambda}&=& -40\sqrt{2}\pi \als {\cal C}_\pi^{(ab)}\frac{C_F}{N_C}
               f_{3\pi} (2\lambda-\mu)\,\frac{\sqrt{-\uh\sh}}{\sh^2\uh^2} \nn\\
               &\times& \Big[ \big(1-\frac3{16} \omega_{10} + \frac6{25} \omega_{20}
                 -\frac3{50}\omega_{11}\big) \big(e_a\uh^2+e_b\sh^2\big)  \nn\\
                 &+& 6 \frac{C_G}{C_F}\big(1-\frac58\omega_{10} + \frac25\omega_{20}
                 + \frac1{10}\omega_{11}\big) \th\big(e_a\uh+e_b\sh\big)   \Big]\,. 
\label{eq:integrated-amp}
  \ea             

  In \ci{KPK18} the following 3-body \da{} has been used to fit the CLAS data \ci{CLAS17}
  on $\pi^0$  photoproduction: 
    \ba
    f_{3\pi}(\mu_0)&=& 0.004\,\gev^2\,,  \nn\\
    \omega_{10}(\mu_0)&=&-2.55\,, \quad \omega_{20}(\mu_0)\=8.0\,, \quad \omega_{11}(\mu_0)\=0\,.
    \label{eq:DA1}
    \ea
   
    The parameter $f_{3\pi}$ is taken from a QCD sum rules analysis and is subject
    to an uncertainty of about $30\%$ \ci{ball99}.
    
    Another 3-body \da{} has been advocated for in \ci{duplancic}:
   \ba
    f_{3\pi}(\mu_0)&=& 0.004\,\gev^2\,,  \nn\\
    \omega_{10}(\mu_0)&=&2.5\,, \quad \omega_{20}(\mu_0)\=6.0\,, \quad \omega_{11}(\mu_0)\=0\,.
    \label{eq:DA2}
    \ea
    This \da{} leads to a fit to the CLAS photoproduction data of similar quality as \req{eq:DA1}
    but, in contrast to \req{eq:DA1},  also to fair agreement with the data on deeply virtual pion
    electroproduction \ci{clas14,hallA20,compass19}. In earlier analyses \ci{GK5,GK6} of deeply
    virtual pion electroproduction the Wandzura-Wilczeck approximation has been used, i.e.\ the
    3-body twist-3 contribution has been neglected.

%%%%%%%%%%%%%%%%%%%%%%%%%%%%%%%%%%%%%%%%%%%%%%%%%%%%%%%%%%%%%%%%%%%%%%%
    \section{The GPDs in the large-$N_C$ limit}
    \label{sec:gpd}
%%%%%%%%%%%%%%%%%%%%%%%%%%%%%%%%%%%%%%%%%%%%%%%%%%%%%%%%%%%%%%%%%%%%%%%%%
As already mentioned there are 14 GPDs contributing to the photoproduction amplitudes and all
of them are unknown at present. In order to achieve estimates of the cross section for the
process of interest  recourse will be taken to large-$N_C$ results. In the limit of large $N_C$
the nucleon and the $\Delta(1232)$ are different excitations of the chiral soliton and they
are degenerated in mass. This leads to relations between proton-$\Delta^+$ and proton-proton
matrix elements of quark-field operators. The following results for the $p-\Delta^{++}$ GPDs,
defined in \ci{belitsky,KPK22} are to be found in the literature~\footnote{
       In \ci{belitsky} the large-$N_C$ relation for $G_1$ is quoted as
       $$  G_1^{p\Delta^{++}}=3/2\Big[H^u-H^d + E^u-E^d\Big]\,. $$
       However, the contribution from $H$ is suppressed by $1/N_C$ compared to $E$ \ci{weiss16}
       and therefore neglected here.}
\ba
G_1^{p\Delta^{++}} &=&\phantom{-}\frac32 \Big[ E^u-E^d\Big]\,,  \hspace*{0.205\tw} \ci{belitsky} \nn\\
\tilde{G}_3^{p\Delta^{++}} &=& - \frac32 \Big[\widetilde{H}^u-\widetilde{H}^d\Big]\,,  \hspace*{0.2\tw}
\ci{frankfurt99} \nn\\
G_{T5}^{p\Delta^{++}} +\frac12 G_{T7}^{p\Delta^{++}} &=& -\frac32\Big[H_T^u-H_T^d\Big]\,.
                                \hspace*{0.205\tw} \ci{KPK22}
\label{eq:large-NC}
\ea
If all other transition GPDs and their associated form factors are neglected there are only four
form factors contributing to wide-angle photoproduction of $\pi^-\Delta^{++}$:
\ba
R_1(t)&=&\phantom{-} \frac32\,\int_0^1 \frac{dx}{x} \Big[E^u(x,t)-E^d(x,t)\Big]\,, \nn\\
\widetilde{R}_3(t)&=& -\frac32 \int_0^1 \frac{dx}{x} \Big[\widetilde{H}^u(x,t)
                            -\widetilde{H}^d(x,t)\Big]\,,\nn\\
S_{T}(t)&=&-\frac32\,\int_0^1 \frac{dx}{x} \Big[H^u_T(x,t)-H^d_T(x,t)\Big]\,,
\ea
According to \req{eq:large-NC} $S_T=S_{T5}+S_{T7}/2$. In order to split $S_T$ in $S_{T5}$
and  $S_{T7}$ a parameter $\varrho$ is introduced such that
\be
S_{T5}\=(1-\varrho) S_T\,, \qquad S_{T7}\=2\varrho S_T\,.
\label{eq:assumption1}
\ee
The parameter $\varrho$ is varied between 0 and 1. This avoids a change of the sign of the form factors.
A possible $t$-dependence of $\varrho$ is ignored. Under these conditions
the process amplitudes \req{eq:amplitudes} read~\footnote{
                   In principle there is also a contribution from the pion pole which can be
                   neglected in the wide-angle region because the pole at $t=m_\pi^2$ is far away
                   ($m_\pi$ denotes the mass of the pion).}
\ba
   {\cal M}_{03,\mu+}&=& \frac{e_0}{\sqrt{2}}\,\Big(\frac{\sqrt{-t}}{2m}
     \sum_\lambda {\cal H}_{0\lambda,\mu\lambda}\,R_1  - {\cal H}_{0-,-\mu +} S_{T7}\Big)\,, \nn\\
         {\cal M}_{01,\mu+}&=& \frac{e_0}{2\sqrt{6}} \frac1{mM}\,   
                \sum_\lambda{\cal H}_{0\lambda,\mu\lambda}\,\Big(t R_1  
               + 2\lambda\,  m(M+m) \widetilde{R}_3 \Big)  \nn\\
         &-& \frac{e_0}{2\sqrt{6}} \frac{\sqrt{-t}}{M}\Big(  
    \sum_\lambda{\cal H}_{0-\lambda,\mu\lambda}\,S_{T5} - 2{\cal H}_{0-,-\mu+}\,S_{T7}\Big)\,,   \nn\\
         {\cal M}_{0-1,\mu+}&=& \frac{e_0}{2\sqrt{6}} \frac{\sqrt{-t}}{M}\,
           \sum_\lambda {\cal H}_{0\lambda,\mu\lambda}\,\Big( R_1 
           - 2\lambda\,\widetilde{R}_3\Big)    \nn\\
         &-& \frac{e_0}{\sqrt{6}} \, \frac{m}{M} {\cal H}_{0-,\mu+} \Big( \frac{M+m}{m} S_{T5}
           + S_{T7}\Big)\,, \nn\\
        {\cal M}_{0-3,\mu+}&=& 0\,.   
         \label{eq:amplitudes-4}
         \ea
  The proton and the $\Delta(1232)$ masses are denoted by $m$ and $M$, respectively.       
 The amplitudes for negative proton helicity follow from from parity invariance
\be
{\cal M}_{0-\nu',-\mu-\nu}\=(-1)^{\mu-\nu+\nu'}\,{\cal M}_{0\nu',\mu\nu}\,.
\ee

\vspace*{0.05\tw}
A flavor symmetric sea is assumed for the proton. Therefore, the isovector combination of a
proton-proton GPD, $K^u_i-K^d_i$,  is equal to the difference of the $u$ and $d$ valence-quark GPDs.
The latter are parameterized as \ci{DFJK4,DK13}
\be
K_i^a(x,t)\=k_i^a(x) \exp{[tf_i^a(x)]}
\label{eq:par}
\ee
for flavor $a$. The profile function reads
\be
 f_i^a(x)\=(B_i^a-\alpha_i'{}^a\ln{x})(1-x)^3 + A_i^ax(1-x)^2\,.
 \label{eq:profile}
 \ee
 This is a generalization to the large $-t$ region of the Regge-like parameterization
 frequently used in the analysis of  deeply virtual exclusive processes. An important property of this 
 profile function is the strong $x-t$ correlation: large (small) $x$ go together with large (small) $-t$.
 It should be noted that the handbag mechanism probes the GPDs \req{eq:par} at large $x$ since it
 applies to large $-t$. In this region where the active parton carries most of the proton's momentum
 while all spectators are soft. This is the region of the Feynman mechanism.
 The parameters of the profile functions for the three relevant proton-proton GPDs are compiled in
 Tab.\ \ref{tab:profile}.
 
 The forward limit of $\widetilde{H}$ is given by the polarized parton densities
 \be
 \tilde{h}^a(x)\=\Delta q^a(x)
 \ee
 which are taken from \ci{dssv}.
 That of $H_T$ is related the transversity distribution. Following an ansatz proposed in
 \ci{anselmino}, $h_T$ is parameterized as \ci{GK6} 
 \be
 h^a_T(x)\=N_T^a\sqrt{x}(1-x)\big[q^a(x)+\Delta q^a(x)\big]\,.
 \ee
 With this parameterization the Soffer bound is respected. The unpolarized parton densities,
 $q^a(x)$, are taken from \ci{pumplin}.  The normalization $N_T^a$ is \ci{GK6}
 \be
 N_T^u\=1.1\,, \qquad N_T^d\=-0.3\,.
 \ee
 In contrast to the $H$-type GPDs the forward limit of $E$ is not accessible in deep inelastic
 electron-proton scattering. Therefore, it is parameterized analogously to the parton densities but
 with new parameters which are fixed in an analysis of the electromagnetic form factors
 of the nucleon performed by \ci{DK13}
 \be
 e^a(x)\=\kappa_aN_ax^{-\alpha_a}(1-x)^{\beta_a} (1+\gamma_a\sqrt{x})\,.
 \label{eq:e}
 \ee
 The normaliztion $N_a$ ensures
 \be
 \int_0^1dx e_a(x)\=\kappa_a\,.
 \ee
 The quantity $\kappa_a$ is the contribution of quarks with flavor $a$ to the anomalous magnetic
 of the proton. Its value is computed from the anomalous magnetic moments of the proton and neutron:
 $\kappa_u=1.67$, $\kappa_d=-2.03$. The other parameters in\req{eq:e} are \ci{DK13}:
 \ba
 \alpha_u&=&0.603\,, \hspace*{0.08\tw} \alpha_d\=0.603\,, \nn\\
 \beta_u&=&4.65\,,    \hspace*{0.095\tw} \beta_d\=5.25\,, \nn\\
 \gamma_u&=&4\,,   \hspace*{0.135\tw} \gamma_d\=0\,.
 \ea
 
 \begin{table*}[t]
\renewcommand{\arraystretch}{1.4} 
\caption{The parameters of the profile functions. Those for the  valence-quark GPD $E$ are taken
  from \ci{DK13}, for $\widetilde{H}$ from \ci{htilde}  and for $H_T$ from \ci{KPK18}.}
\begin{center}
\begin{tabular}{| c || c  c | c  c | c  c|}
                    & $E^u$&  $E^d$& $\widetilde{H}^u$& $\widetilde{H}^d$&$H_T^u$ & $H_T^d$ \\[0.2em]
  \hline
  $\alpha' [\gev^{-2}]$& 0.961 & 0.861& 0.432 & 0.387& 0.45 & 0.45 \\[0.2em]
  $B [\gev^{-2}]$& 0.333 & -0.635& 0.654 & 0.400 & 0.3 & 0.3 \\[0.2em]
  $A  [\gev^{-2}]$& 1.187 & 3.106& 1.239 & 4.284 & 0.5 & 0.5 \\[0.2em]
 \hline
\end{tabular}
\end{center}
\label{tab:profile}
\renewcommand{\arraystretch}{1.0}
 \end{table*}

\begin{figure}[th]
  \begin{center}
 \includegraphics[width=0.4\tw]{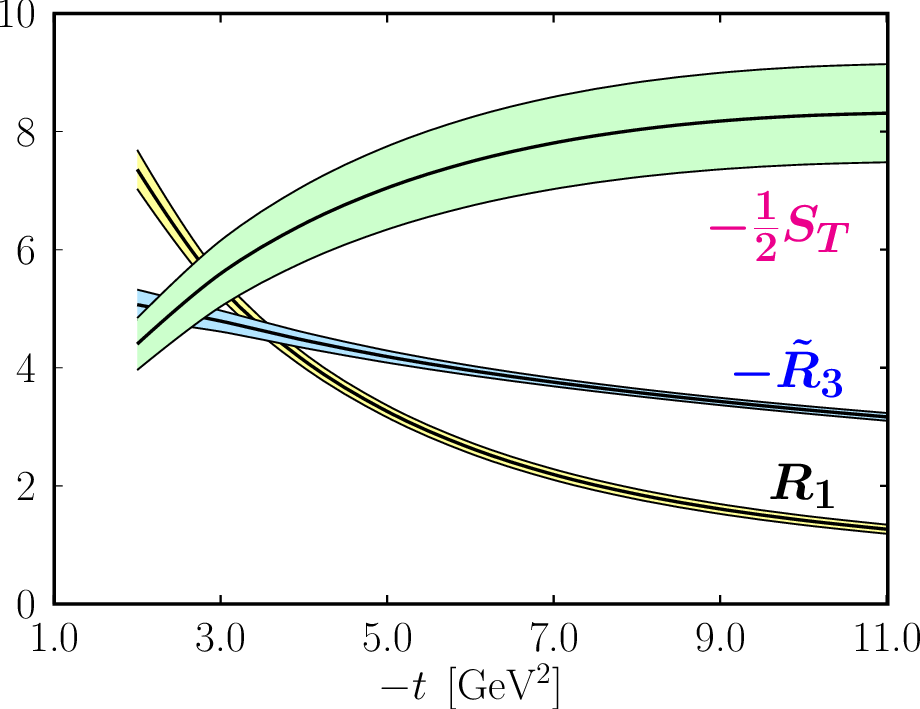}\hspace*{0.05\tw}
 \includegraphics[width=0.4\tw]{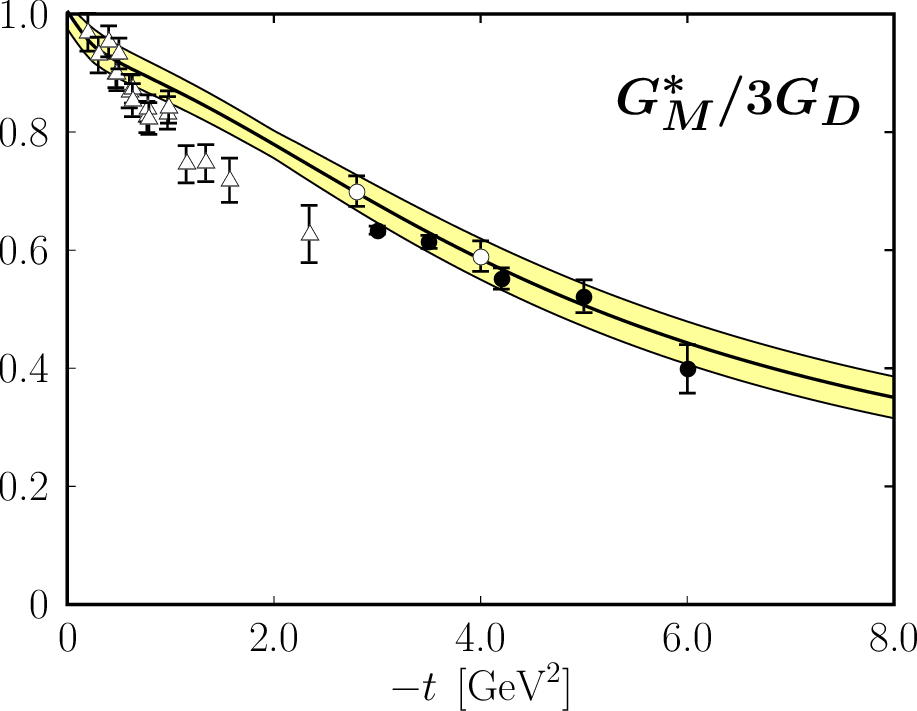}
  \end{center}
  \caption{(Left) The proton-$\Delta^{++}$ form factors in the large-$N_C$ limit scaled
    by $t^2$ versus $-t$. The shaded bands indicate the uncertainties of the corresponding
    proton-proton form factors.}{\label{fig:FA}}
  \vspace*{-0.025\tw}\caption{(Right) The magnetic $p-\Delta^+$ transition form factor $G_M^*$
    scaled by three times the dipole form factor $G_D=(1-t/0.71\,\gev^2)^{-2}$ versus $-t$. The
    shaded band represents the uncertainty of the prediction. Date are taken from
    \ci{bartel68,frolow99,CLAS06}.}{\label{fig:GM}}
 \end{figure}

 In Fig.\ \ref{fig:FA} the $p-\Delta^{++}$ form factors are displayed. The error bands are those of
 the corresponding proton-proton form factors, see \ci{KPK18,DK13}. 

 The mentioned zero-skewness proton-proton GPDs have been extracted and used in analyses of the
 electromagnetic and axial form factors of the nucleon, wide-angle Compton scattering and pion
 photoproduction. With the skewness dependence generated through double distributions \ci{musatov}
 they have also been applied in analyses of deeply virtual exclusive processes. Another test of the
 GPDs is provided by the magnetic $p-\Delta^*$ transition form factor, $G_M^*$, for which data
 are available at fairly large values of $-t$ \ci{bartel68,frolow99,CLAS06}, see Fig.\ \ref{fig:GM}.
 The experimental value of $G_M^*(0)$ has been extracted from the MAMI data \ci{MAMI00} on the
 $\gamma N\Delta$ amplitudes at the resonance position \ci{tiator01}:
 \be  
 G_M^*(0)\=3.02\pm 0.03\,.
\label{eq:GM-exp}
 \ee
In the large-$N_C$ limit $G_M^*$ is related to the proton-proton GPD $E$ by \ci{belitsky}
\ba
G_M^*(t)&=& -\frac13\int_0^1 dx G_1^{p\Delta^+}(x,t)\=\frac1{\sqrt{3}} \int_0^1dx \Big[E^u(x,t)-E^d(x,t)\Big]
                                          \nn\\
                                          &=& \frac1{\sqrt{3}} \Big[F_2^u(t)-F_2^d(t) \Big]
\label{eq:GM}                                          
\ea
where $F_2^a$ denotes the flavor-$a$ Pauli form factor of the nucleon. At $t=0$ in particular
one obtains from \req{eq:GM}
\be
G_M^*(0)\= \frac1{\sqrt{3}} \big(\kappa_u-\kappa_d\big) \= 2.14\,.
\ee
This value of $G_M^*(0)$ is too small by about $30\%$. Errors of this size are to be expected
for large $N_C$ results~\footnote{
        Taking into account corrections suppressed by $1/N_C$ compared to \req{eq:GM},
        one obtains $G_M^*(0)=2.71$ and for the magnetic moment $\mu_{p\Delta^+}=(\mu_p-\mu_n)/\sqrt{2}$
        \ci{belitsky,jenkins94}.}. 
In order to test the $t$-dependence of the large-$N_C$ prediction, we follow \ci{pascalutsa}
and  normalize the magnetic transition form factor $G_M^*$ to the experimental
value \req{eq:GM-exp}:
\be
G_M^*(t)\=\frac{3.02}{\kappa_u-\kappa_d} \Big[F_2^u(t)-F_2^d(t)\Big]
\ee
In Fig.\ \ref{fig:GM} the experimental data on the magnetic $p-\Delta^+$ form factor $G_M^*$
are compared
with the large-$N_C$ predictions computed from the GPD $E$ respective the flavor Pauli form factors
proposed in \ci{DK13}. For $-t$ less than $2\,\gev^2$ there are some discrepancies with the data of
\ci{bartel68}. On the other hand, for $-t$ larger than about $2\,\gev^2$
- this is the region relevant in this work - there is good agreement with experiment, i.e.
the $t$-dependence of $G_M^*$ is well in agreement with the large-$N_C$ prediction.
That the large-$N_C$ relation between the GPDs $G_1$ and $E$ (see Eq. \req{eq:large-NC})
implies similar shapes of the Pauli form factor and $G_M^*$ has already been noticed by Stoler \ci{stoler}.

The parameterization \req{eq:par} of the GPDs with the profile function \req{eq:profile}
has a remarkable property: As shown in \ci{DFJK4} the moments of the GPDs
behave power-law like at large $-t$. The power is controlled by the power $\beta_i$ of
the factor $1-x$ that characterizes the behavior of a GPD $K_i$ for $x\to 1$. Thus, for instance
\be
F_2^a \sim (-t)^{-d_{e}^a}
\ee
with $d_{e}^a=(1+\beta_{e}^a)/2$. For the Pauli form factor the powers are \ci{DK13}
\be
d_{e}^u\=2.83\,, \qquad d_{e}^d\=3.12\,.
\ee
As is to be seen from Fig.\ \ref{fig:GM} $G_M^*$ falls faster than $1/t^2$ and seems to be
in agreement with the dominance of the  GPD $E$. Since the valence quark densities behave about as
$(1-x)^{3.4}$ for $x\to 1$ for $u$-quarks \ci{DK13} and a somewhat stronger fall-off for $d$-quarks, the
$u$-quark contribution dominates the form factors $\widetilde{R}_3, S_{T5}$ and $S_{T7}$ at
large $-t$. Hence, the form factor $\widetilde{R}_3$ falls approximately as $(-t)^{-2.2}$
and $S_{T5}$, $S_{T7}$ as $(-t)^{-2.7}$. This power-law behavior of the GPD-moments obtained with the
profile function \req{eq:profile} is to be contrasted with the familiar Regge-like profile
function, i.e.\ $A_i=0$ in \req{eq:profile}, which leads to  exponentially decreasing GPD moments.
The powers $\beta_i$ extracted from parton densities or from electromagnetic form factors \ci{DK13}
are to be considered as effective powers since they are fixed in regions of $x$ less than about
$0.8$. They are likely subject to change as soon as data sensitive to larger $x$ become available.
It is well possible that at last the powers agree with the theoretical expectations for $x\to 1$ \ci{yuan},
e.g. ($a=u,d$)
\be
H^a \sim (1-x)^3\,, \qquad E^a\sim (1-x)^5\,.
\ee

%%%%%%%%%%%%%%%%%%%%%%%%%%%%%%%%%%%%%%%%%%%%%%%%%%%%%%%%%%%%%%%%%%%%%%
\section{Observables, predictions and comparison \\ with experiment}
%%%%%%%%%%%%%%%%%%%%%%%%%%%%%%%%%%%%%%%%%%%%%%%%%%%%%%%%%%%%%%%%%%%%%%
Now, having specified the \da s and the form factors, we are in the position to compute the amplitudes
\req{eq:amplitudes-4} and subsequently the differential cross section for $\pi\Delta$ photoproduction
\be
\frac{d\sigma}{dt}\=\frac1{32\pi (s-m^2)^2}\sum_{\nu'\mu} |{\cal M}_{0\nu',\mu +}|^2\,.
\ee

\begin{figure}[t]
  \begin{center}
    \includegraphics[width=0.395\tw]{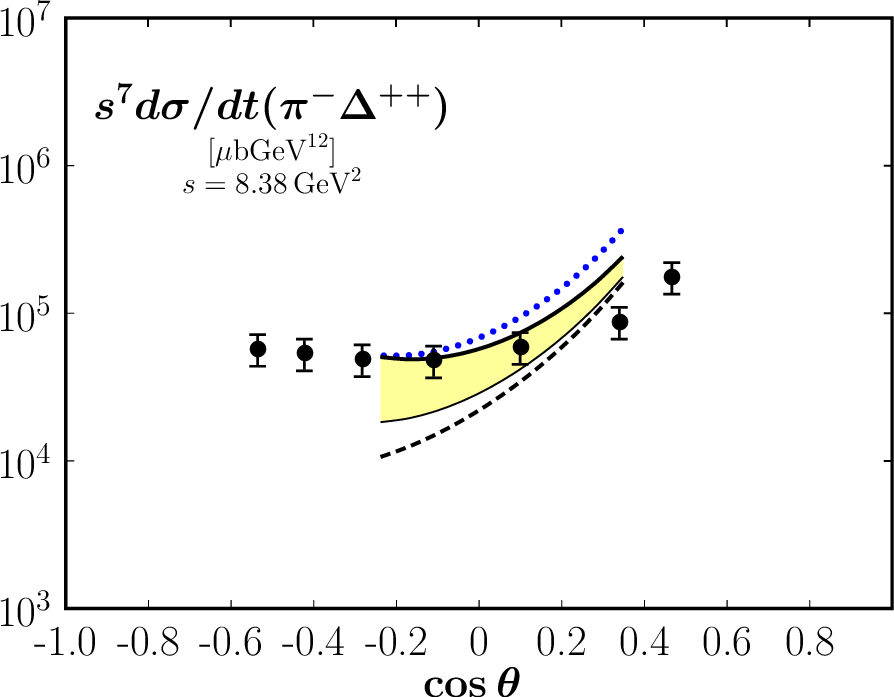}\hspace*{0.03\tw}
    \includegraphics[width=0.395\tw]{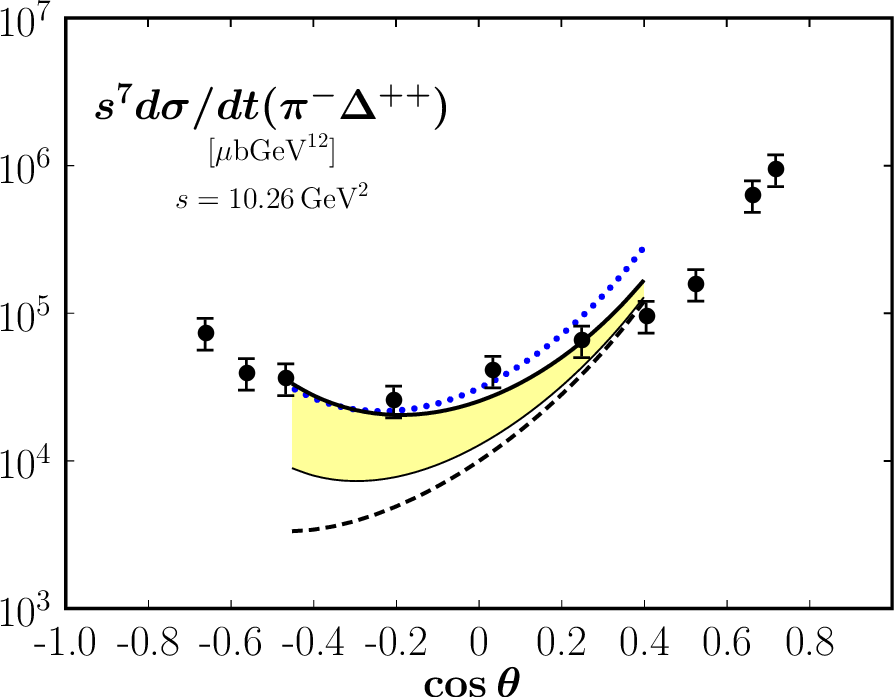}
\end{center}
  \caption{(Left) The differential cross section for $\pi^-\Delta^{++}$, scaled by $s^7$, at $s=8.38\,\gev^2$
    versus $\cos{\theta}$. The solid (dashed, dotted) line is evaluated from $\varrho=1.0$ and the
    \da{} \req{eq:DA2} (the twist-2 contribution, $\pi^+\Delta^0$). The shaded band represents the
    dependence on $\varrho$ varying it between 0 and 1. Theoretical results are only shown for $-t$ and
    $-u$ larger than $2.5\,\gev^2$. Data is taken from \ci{anderson}.}{\label{fig:dsdt-1}}
  \vspace*{-0.025\tw}\caption{(Right) As Fig.\ \ref{fig:dsdt-1} but for $s=10.26\,\gev^2$.\hspace*{0.3\tw}}
    {\label{fig:dsdt-2}}   
\end{figure}
In Figs.\ \ref{fig:dsdt-1} and \ref{fig:dsdt-2} the predictions for the $\pi^-\Delta^{++}$ cross section,
scaled by $s^7$, are shown at two values of $s$ and compared with the available data \ci{anderson}. The
cross section is displayed versus the cosine of the c.m.s. scattering angle $\theta$.
The predictions are evaluated from the 3-body \da{} \req{eq:DA2} and $\varrho$ is fitted to the data. Fair
agreement with experiment is obtained for $\varrho=1.0$ (i.e.\ $S_{T5}=0$, $S_{T7}=2S_T$). It is to be
stressed that this is the only parameter that is fitted to the data. All other input functions, \da s and
GPDs, are fixed in studies of deeply virtual and wide-angle processes with exclusive photon- and
meson-nucleon final states \ci{KPK18,KPK21,duplancic,GK5,GK6,htilde,GK3} as well as in the analysis of
the electromagnetic form factors \ci{DK13}. The dependence of the predicted cross section on the value
of $\varrho$ is shown as a shaded band in Fig.\ \ref{fig:dsdt-1}. This dependence constitutes a substantial
part of the uncertainties of the predictions. For the parameter $A_T^u=A_T^d$ in the profile function
of the GPD $H_T$ the value $0.5\,\gev^{-2}$ is chosen as in \ci{KPK18,KPK21}. It is not well constrained
by the available data on photoproduction of $\pi N$ and $\pi\Delta$ states \ci{anderson,CLAS17}, it can
be varied between 0.3 and $0.7\,\gev^{-2}$. It should also be mentioned that the theoretical results are
only shown for $-t$ and $-u$ larger than $2.5\,\gev^2$ in order to meet approximately the requirement of
the handbag mechanism that the Mandelstam variables should be (much) larger than a typical hadronic scale
of order $1\,\gev^2$. Baryon-mass corrections have been studied in \ci{DFHK} by testing other possibilities
to assign the subprocess Mandelstam variables to the full ones than \req{eq:assignment}.
According to \ci{DFHK} the mass corrections are not small at $s\simeq 10\,\gev^2$.

In the figures the twist-2 contribution is separately shown. It dominates in the forward hemisphere
whereas in the backward region the twist-3 contributions takes the lead. This is also the case for
photoproduction of charged pions and nucleons \ci{KPK21}. In contrast to the latter processes
the $\pi^+\Delta^0$ cross section is larger
than the $\pi^-\Delta^{++}$ one in the forward hemisphere (at $s=10.26\,\gev^2$ about a factor of
1.7 at $\cos{\theta}\simeq 0.4$, slightly increasing with energy) but becomes a little bit smaller
in the backward hemisphere (about a factor of 0.95 at $\cos{\theta}\simeq -0.4$) although the
subprocess amplitudes for $\pi^-$ production are larger than those for $\pi^+$ (see Eq.\ \req{eq:twist3})
\be
\frac{{\cal H}^{\pi^-}}{{\cal H}^{\pi^+}} \sim \left|\frac{e_d\uh^n+e_u\sh^n}{e_u\uh^n+e_d\sh^n}\right|
\ee
($n=1,2$). The larger $\pi^+$ cross section is caused by the factor $\sqrt{3}$ in the isospin
relation \req{eq:isospin}.
\begin{figure}[t]
  \begin{center}
  \includegraphics[width=0.5\tw]{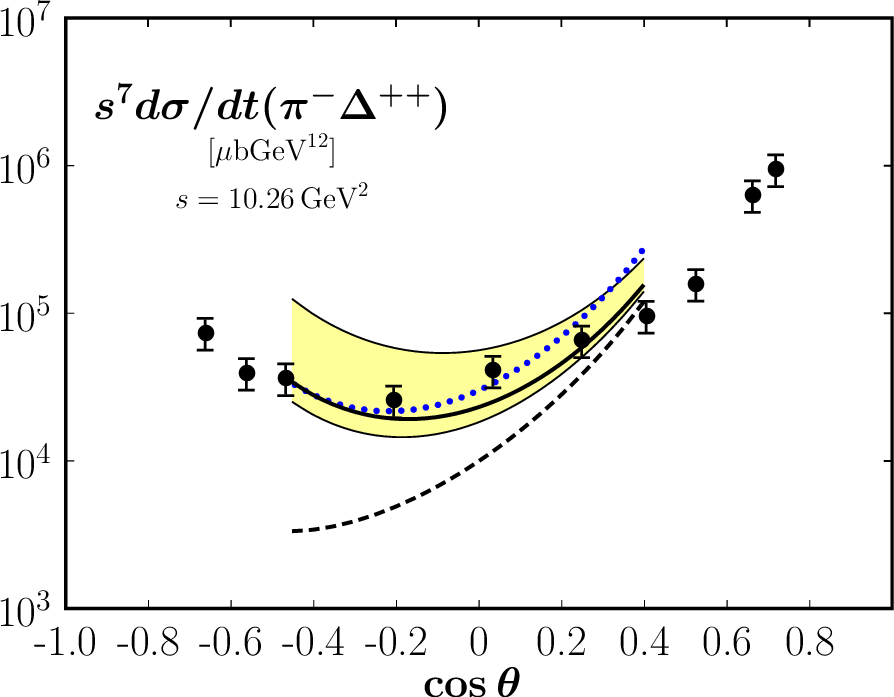}  \hspace*{0.03\tw}
\end{center}
  \caption{The scaled $\pi^-\Delta^{++}$ differential cross section at $s=10.26\,\gev^2$.
    The solid (dashed, dotted) line is evaluated from the \da{} \req{eq:DA1} 
    and the assumption $\varrho=0.2$ (the twist-2 contribution, $\pi^+\Delta^0$).
    For other notations it is referred to Fig.\ \ref{fig:dsdt-1}.}
\label{fig:dsdt-3}
\end{figure}

In Fig.\ \ref{fig:dsdt-3} the cross section evaluated from the \da{}
\req{eq:DA1} together with $\varrho=0.2$ (in this case $S_{T5}=2 S_{T7}$) is shown at
$s=10.26\,\gev^2$. The properties of the predictions evaluated from this \da{} are very similar
to the scenario  discussed above. Also their energy dependency is similar.
It is to be stressed that the \da s \req{eq:DA1}  and \req{eq:DA2} lead to similar results for
$\gamma p\to \pi^0 p$ in the wide-angle region but to very different results for deeply virtual
pion electroproduction \ci{duplancic}. The reason for that difference lies in the fact that to
deeply virtual pion electroproduction there is also a strong contribution from the 2-body twist-2
\da{}, $\phi_{\pi p}$, which is fixed by the 3-body twist-3 \da{} via the equation of motion. The
3-body \da{} \req{eq:3-body-DA} leads to
\be
\phi_{\pi p}(\tau)\=1+\frac{f_{3\pi}}{f_\pi\mu_\pi} \omega (1-30\tau^2\taub^2)
\ee
where
\be
\omega\=7\omega_{10} - 2\omega_{20} -\omega_{11}\,.
\ee
From the \da{} \req{eq:DA1} one finds $\omega(\mu_0)=-33.85$ whereas \req{eq:DA2} leads to
$\omega(\mu_0)=5.5$. The different values of $\omega$ have a strong impact on the results for
electroproduction.

A comparison of the cross sections at $s=8.38$ and $10.26\,\gev^2$ reveals that the handbag results
do not scale as $s^{-7}$. This would only happen for a dominant twist-2 contribution and
if the form factor $R_1$ would drop as $1/t^3$ and $\widetilde{R}_3$ as $(-t)^{-2.5}$  which is not
the case, see the discussion at the end of Sect. \ref{sec:gpd}.  The twist-3 contribution scales
as $s^{-8}$ (if $S_T\sim (-t)^{-2.5}$). Moreover, there is a number of logarithms generated by
the evolution of the distribution amplitudes and GPDs. Thus, effectively the handbag result for the
cross section scales about as $s^{-9.5}$. A possible evolution of the form factors which -
as said - represent $1/x$-moments of
GPDs, is ignored. This is to some extent justified: At large $-t$ the form factors are accumulated
in a narrow range of large $x$, see \req{eq:par}, \req{eq:profile}. Because of the strong $x-t$
correlation of the GPDs this range of $x$ approaches 1 and becomes narrower for increasing $-t$.
Therefore, for very large $-t$ the form factors approximately become equal to the scale-independent
lowest moments of the GPDs. One may argue that the disregard of the GPD evolution also requires
the neglect of the scale-dependence of the \da s for consistency. Thus, as an example, the cross
section is evaluated from the \da{} \req{eq:DA2} taken at the fixed scale $\muF=\muR=1\,\gev$ and with
$\varrho=0.4$. The result agrees also reasonable well with experiment but, as expected, shows a milder
energy dependence. Lack of data prevent a serious check of the energy dependence.

\begin{figure}[t]
  \begin{center}
    \includegraphics[width=0.4\tw]{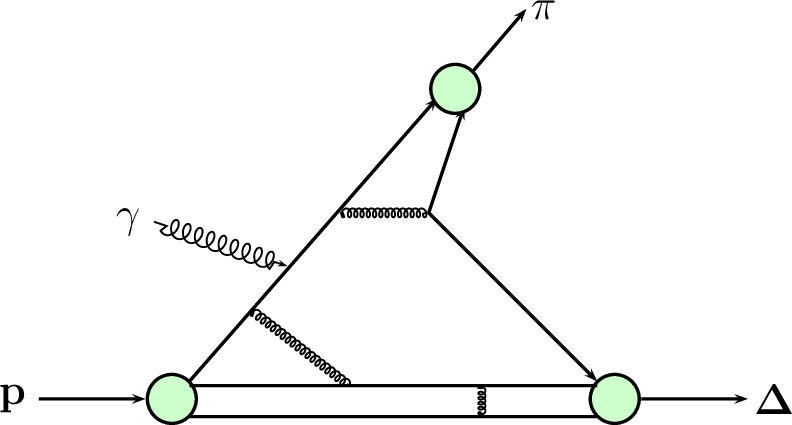}
\end{center}
  \caption{A typical graph for the hard perturbative mechanism \ci{brodsky80,efremov80}.
  The shaded regions are soft regions in which no hard scale occur.}
  \label{fig:BL-graph}
  \end{figure}
There is an alternative mechanism for wide-angle scattering \ci{brodsky80,efremov80} for which,
in contrast to the handbag mechanism, all partons inside the baryons participate in the hard
process by the exchange of hard gluons. The dominant contribution in this case comes from the valence
Fock component where the number of exchanged gluons is minimal, see Fig. \ref{fig:BL-graph}.
In the handbag mechanism, on the other hand, there is only one active parton, all others are
spectators and it is summed over all Fock states, see Fig. \ref{fig:handbag}.
However, a calculation of $\gamma p\to \pi^-\Delta^{++}$ in the hard perturbative mechanism
fails by about three orders of magnitude at $\theta\simeq 90^\circ$ \ci{farrar91}.

Data on spin-dependent observables would give detailed information on the various $p-\Delta$ transition
GPDs. Such observables could be measured with polarized photons and protons. The spin density matrix
elements of the $\Delta(1232)$ would provide further information on the GPDs. At present just four out
14 GPDs are fixed by the large-$N_C$ relations \req{eq:large-NC}, the others are assumed to be zero.
This scenario is probably inadequate to explain all spin phenomena. Thus, likely, the large-$N_C$ GPDs
 \req{eq:large-NC} do not allow reliable predictions
of spin-dependent observables. An exception is perhaps the correlation of the photon and
proton helicities, $A_{LL}$, which like the cross section, only depends on the absolute values of
the amplitudes:
\ba
A_{LL}&=& \frac{d\sigma(++) - d\sigma(-+)}{d\sigma(++) + d\sigma(-+)} \nn\\
&=& \frac{\sum_{\nu'}\Big[ |\Phi_{0\nu',++}|^2-|\Phi_{0\nu',-+}|^2\Big]}
         {\sum_{\nu'}\Big[ |\Phi_{0\nu',++}|^2+|\Phi_{0\nu',-+}|^2\Big]}
\ea
where $d\sigma(\mu\nu)$ is the cross section with definite initial state helicities
The amplitudes $\Phi$ are standard c.m.s. helicity amplitudes which are more convenient
for the treatment of spin-dependent observables than the light-cone helicity amplitudes. The
amplitudes $\Phi$ can be obtained from the light-cone helicity amplitudes ${\cal M}$
\req{eq:amplitudes-4} which
naturally appear in the handbag mechanism, by using the transform from light-cone spinors to
ordinary spinors  given in \ci{diehl01}. This transform generates corrections to a given
helicity amplitude $\Phi_{0\nu',\mu+}\approx {\cal M}_{0\nu',\mu+}$ of order
\be
\eta_0\=\frac{2\sqrt{-t}}{\sqrt{s}+\sqrt{-u}}\,\frac{M (m)}{\sqrt{s}}
\ee
from other light-cone helicity amplitudes.

\begin{figure}[t]
  \begin{center}
  \includegraphics[width=0.5\tw]{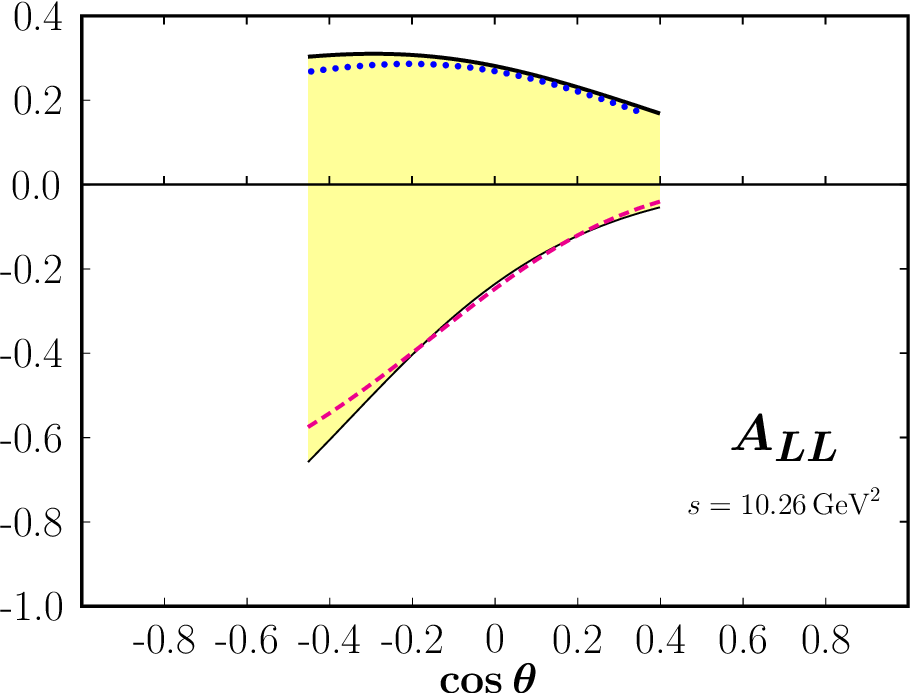}\hspace*{0.03\tw}
\end{center}
  \caption{The helicity correlation parameters $A_{LL}$ at $s=10.26\,\gev^2$.
    The solid (dotted ) line is evaluated from the \da{} \req{eq:DA2} and  $\varrho=1.0$
    for $\pi^-\Delta^{++}$ ($\pi^+\Delta^0$). The dashed line is computed from the \da{} \req{eq:DA1}
    and $\varrho=0.2$. For other notations it is referred to Fig.\ \ref{fig:dsdt-1}.}
    \label{fig:ALL}
\end{figure}
Predictions of $A_{LL}$ are displayed in Fig.\ \ref{fig:ALL}. The results for the $\pi^-\Delta^{++}$ and
$\pi^+\Delta^0$ channels are evaluated from the \da{} \req{eq:DA2} and $\varrho=1.0$. They are very
similar, about 0.2. The \da{} \req{eq:DA1} (with $\varrho=0.2$) leads to a negative $A_{LL}$.

%%%%%%%%%%%%%%%%%%%%%%%%%%%%%%%%%%%%%%%%%%%%%%%%%%%%%%%%%%%%%%%%%%%%%
\section{Summary}
%%%%%%%%%%%%%%%%%%%%%%%%%%%%%%%%%%%%%%%%%%%%%%%%%%%%%%%%%%%%%%%%%%%%%%
Wide-angle photoproduction of $\pi\Delta(1232)$ final states at large energies is calculated within
the handbag mechanism in which the process amplitudes factorize in hard partonic subprocess amplitudes
and soft form factors representing $1/x$-moments of zero-skewness $p-\Delta$ transition GPDs.
The subprocess $\gamma q_a\to \pi q_b$ is the same as in pion photoproduction. The corresponding
amplitudes have been calculated in \ci{KPK18} to twist-3 accuracy and to leading-order
of perturbative QCD. The results given in \ci{KPK18} are used here.
The $p-\Delta$ transition GPDs are related to the proton-proton GPDs in the large-$N_C$
limit. The latter GPDs are known from analyses of the electromagnetic form factors, wide-angle Compton
scattering and pion photoproduction as well as from investigations of deeply virtual meson production.
There is only one issue that implies an unknown parameter: In the large-$N_C$ limit only the sum of
transversity GPDs $S_{T5}+1/2S_{T7}$ is related to a proton-proton GPD. The splitting of the relation
\req{eq:large-NC} into $S_{T5}$ and $S_{T7}$ is done with the help of a parameter $\varrho$
(see \req{eq:assumption1}) which is fitted to the experimental data on $\gamma p\to \pi^-\Delta^{++}$
\ci{anderson}. Reasonable agreement with experiment is achieved that way for two different 3-body
twist-3 \da s. These \da s also lead to similar results for pion photoproduction but to very different
results for deeply virtual electroproduction of pions \ci{duplancic}. A remeasurement of the
cross section seems to be advisable since the SLAC data \ci{anderson} are very old. Data at higher energies
as well as spin-dependent data would be welcome. The handbag mechanism also applies to other photoproduction
channels like $\eta\Delta^+$, $\rho\Delta$ or kaon-$\Sigma^*$ but there are no data available for these
processes at present.

%%%%%%%%%%%%%%%%%%%%%%%%%%%%%%%%%%%%%%%%%%%%%%%%%%%%%%%%%%%%%%%%%%%%%%%%%%%%%%%

\end{document}